\documentclass[pre,onecolumn,amssymb]{revtex4}
\usepackage{epsfig}
\usepackage{amssymb}

\begin{document}
\title{Computational and Analytical Modeling of Cationic Lipid-DNA Complexes}
\author{Oded Farago}
 \affiliation{Department of Biomedical Engineering, Ben Gurion
 University, Be'er Sheva 84105, Israel} 
\author{Niels Gr{\o}nbech-Jensen}
 \affiliation{Department of Applied Science, University of California,
 Davis, California 95616}
\date{\today}

\begin{abstract}

We present a theoretical study of the physical properties of cationic
lipid-DNA (CL-DNA) complexes - a promising synthetically based
nonviral carrier of DNA for gene therapy. The study is based on a
coarse-grained molecular model, which is used in Monte Carlo (MC)
simulations of mesoscopically large systems over time scales long
enough to address experimental reality. In the present work we focus
on the statistical-mechanical behavior of lamellar complexes, which in
MC simulations self-assemble spontaneously from a disordered random
initial state. We measure the DNA-interaxial spacing, $d_{\rm DNA}$,
and the local cationic area charge density, $\sigma_M$, for a wide
range of values of the parameter $\phi_c$ representing the fraction of
cationic lipids. For weakly charged complexes (low values of
$\phi_c$), we find that $d_{\rm DNA}$ has a linear dependence on
$\phi_c^{-1}$, which is in excellent agreement with x-ray diffraction
experimental data. We also observe, in qualitative agreement with
previous Poisson-Boltzmann calculations of the system, large
fluctuations in the local area charge density with a pronounced
minimum of $\sigma_M$ halfway between adjacent DNA molecules.  For
highly-charged complexes (large $\phi_c$), we find moderate charge
density fluctuations and observe deviations from linear dependence of
$d_{\rm DNA}$ on $\phi_c^{-1}$. This last result, together with other
findings such as the decrease in the effective stretching modulus of
the complex and the increased rate at which pores are formed in the
complex membranes, are indicative of the gradual loss of mechanical
stability of the complex which occurs when $\phi_c$ becomes large. We
suggest that this may be the origin of the recently observed enhanced
transfection efficiency of lamellar CL-DNA complexes at high charge
densities, because the completion of the transfection process requires
the disassembly of the complex and the release of the DNA into the
cytoplasm. Some of the structural properties of the system are also
predicted by a continuum free energy minimization. The analysis, which
semi-quantitatively agrees with the computational results, shows that
that mesoscale physical behavior of CL-DNA complexes is governed by an
interplay between electrostatic, elastic, and mixing free energies.

\end{abstract}
\maketitle

\section{Introduction}
\label{intro}

Somatic gene therapy holds great promise for future medical
applications, for example, as new treatment for various inherited
diseases and cancers \cite{therapy1,therapy2}. Viral vectors have been
the most widely used systems for this purpose \cite{viral1,viral2},
but synthetic nonviral vectors are emerging as an attractive
alternative because of their inherent advantages
\cite{synthetic1,synthetic2,synthetic3}. These advantages include ease and
variable preparation, unlimited length of the transported DNA, and
lack of specific immune response due to the absence of viral peptide
and proteins \cite{synthetic3,synthetic4,synthetic5}. Complexes
consisting of cationic lipids (CLs) and DNA comprise one of the most
promising classes of nonviral vectors. They are already used widely
for {\em in vitro}\/ transfection of mammalian cells in research
applications, and have even reached the stage of empirical clinical
trials \cite{trials}. Currently, their efficiency of gene transfer is
considerably lower than that of viral vectors
\cite{efficiency1,safinya_currmed}. Substantial improvement of their
efficiency is required before CL-DNA complexes become available for
therapeutic purposes.

CL-DNA complexes are formed spontaneously when DNA is mixed with
cationic and natural lipids in an aqueous environment
\cite{safinya_science1,safinya_science2}. Their formation is driven by
the electrostatic attraction between negatively charged DNA and
cationic lipid headgroups, and through the entropic gain associated
with the concurrent release of the tightly bounded counterions from
the CL and DNA
\cite{safinya_science1,safinya_science2,harries_may_benshaul,harries_may_benshaul1}. X-ray
diffraction experiments have revealed that CL-DNA complexes exist in a
variety of mesoscopic structures
\cite{lasic,safinya_currop}. These structures include a multilamellar
phase where DNA monolayers are intercalated between lipid bilayers
(${\rm L_{\alpha}^{\ C}}$)
\cite{safinya_science1}, and an inverted hexagonal phase with DNA
encapsulated within cationic lipid monolayers tubes and arranged on a
two-dimensional hexagonal lattice (${\rm H_{II}^{\ C}}$)
\cite{safinya_science2}. In the more commonly observed ${\rm
L_{\alpha}^{\ C}}$ phase, the DNA chains form a one-dimensional
lattice, where the interaxial spacing $d_{\rm DNA}$ decreases with the
charge density of the membrane. Isoelectric complexes, where the
charges on the DNA exactly match those on the CL, are the most stable
ones since they enable nearly complete counterion release
\cite{harries_may_benshaul}. For transfection, positively charged
complexes are used, which can adhere to the negatively charged cell
plasma membrane \cite{safinya_currmed}.

Despite all the promise of CL-DNA complexes as gene vectors, their
transfection efficiency (TE; the ability to transfer DNA into cells
followed by expression) remains substantially lower than that of viral
vectors\cite{efficiency1,safinya_currmed}. This has spurred an intense
research activity aimed at enhancing TE
\cite{efficiency1,safinya_currmed,te1,te2,te3,te4}. Recent tragic
events associated with the use of engineered adenovirus vectors have
further stimulated the search for efficient synthetic DNA carriers
\cite{tragedy1,tragedy2}. Recognizing that the structure of CL-DNA
complexes may strongly influence their function and TE, much of the
effort in theoretical and experimental studies has been devoted to
understanding the mechanisms governing complex formation, structure,
and phase behavior \cite{safinya_currop,may_benshaul}. The most widely
used approach to describe the free energy of the complexes is the
mean-field Poisson-Boltzmann equation which takes into account the
electrostatic interactions of a charged continuum and the mixing
entropy of the lipids and counterions
\cite{harries_may_benshaul,bruinsma}. The bending energy of the lipid
layer is introduced by means of the effective Helfrich energy
\cite{harries_may_benshaul1,harries_may_benshaul2}. Thus, the
preferred structure is determined by a delicate interplay between
electrostatic interactions and bending elasticity, both depending on
the molecular nature and composition of the lipids. The vast majority
of neutral lipids, when complexed with CLs, lead to the ${\rm
L_{\alpha}^{\ C}}$ phase. Addition of the neutral lipid DOPE (or other
PE-based lipids) drives the spontaneous curvature negative, thereby
inducing the transition to the ${\rm H_{II}^{\ C}}$ phase
\cite{safinya_science2}. This transition is also promoted by addition
of the cosurfactant hexanol which lowers the bilayer bending rigidity.

The relationship between the physical properties and TE of CL-DNA
complexes has been studied in a recent set of experiments utilizing a
combination of several techniques (synchrotron X-ray diffraction for
structure determination, laser scanning confocal microscopy to probe
the interactions of complexes with cells, and luciferase reporter-gene
expression assays to measure TE) \cite{safinya_currmed}. The most
notable result of these experiments is the identification of the
membrane charge density, $\sigma_M$, as a key universal parameter that
governs TE of ${\rm L_{\alpha}^{\ C}}$ complexes
\cite{safinya_charge}. The highest transfection rate has been observed
at intermediate $\sigma_M$, reaching values that are comparable to the
high, $\sigma_M$-independent, TE of DOPE-containing ${\rm H_{II}^{\
C}}$ complexes \cite{safinya_charge2}. That lamellar complexes compete
the high TE of hexagonal complexes is of prime importance because, as
mentioned above, most commonly used lipids prefer the ${\rm
L_{\alpha}^{\ C}}$ over the ${\rm H_{II}^{\ C}}$ phase. Moreover, with
newly synthesized multivalent lipids with headgroups whose charge is
as large as $Z=5$, it is possible to reach the optimal TE with a fewer
number of CLs. This is a desirable feature which reduces the cost and,
more importantly, the toxic effects of the CLs. It also means a
smaller metabolic effort for the elimination of the lipids from the
cell.

Based on the above TE data, as well as X-ray diffraction and laser
scanning confocal microscopy imaging, a model of cellular entry of
${\rm L_{\alpha}^{\ C}}$ complexes has been proposed which suggests
that the process of transfection involves two stages
\cite{safinya_currmed}: (1) cellular uptake via endocytosis, and (2)
escape of the complex from the endosome, presumably through the fusion
of the complex with the endosomal membrane and release of the DNA into
the cytoplasm. The adhesion of the complex to the cell is mediated by
electrostatic attraction between the positively charged complex and
the negatively charged cell's plasma membrane. The transfection process is
limited by the rate of the second step, which increases exponentially
with $\sigma_M$. The independence of ${\rm H_{ II}^{\ C}}$ complexes
TE on $\sigma_M$ (in the low $\sigma_M$ regime) has been attributed to
the mismatch between the {\em positive}\/ curvature of the outermost
lipid monolayer (which provides the complex with hydrophobic
shielding) and the complex {\em negative}\/ spontaneous
curvature. This elastically frustrated state drives, independently of
$\sigma_M$, the rapid fusion of the ${\rm H_{II}^{\ C}}$ complex with
the plasma or endosomal membrane. Lamellar complexes with very high
$\sigma_M$ also exhibit reduced TE which should be attributed to the
inability of the DNA to dissociate from the highly charged membranes
(of the free complex) and to become available for expression
\cite{safinya_charge2}.

Theoretical modeling of large molecular assemblies pose significant
challenges due to the spatial complexity of such systems and by the
range of temporal scales involved. In the case of CL-DNA complexes,
the size of the complex may be as large as $1\mu m$, while the basic
unit cell of the ${\rm L_{\alpha}^{\ C}}$ complex is in the nanometer
range (the DNA spacing is typically $d_{\rm DNA}\sim 20-70$\AA, and
the inter-layer spacing $d\simeq 65$\AA). Since both short- (steric,
hydrophobic), and long-range (electrostatic) interactions determine
the physical and biomedical properties of CL-DNA complexes, it is
essential that these systems will be studied at all possible levels of
detail. Moreover, phase transitions of CL-DNA complexes as well as
other topological changes (e.g., membrane fusion which occurs during
transfection) involves the collective motion of many lipid molecules
and, therefore, inherently take place on a variety of spatial and
temporal scales.

To address the multi-scale nature of CL-DNA complexes, a variety of
models, differing in the length and the time scales of the phenomena
of interest have been devised. At the microscopic molecular level, we
have atomistic molecular dynamics (MD) simulations in which the
lipids, DNA, and the embedding solvent are modeled explicitly in full
(classical) atomic detail
\cite{bandyopadhyay_terek_klein}. These simulations provide valuable
information regarding the molecular structure of the complexes, such
as the role played by the neutral PC headgroups (more specifically,
the ${\rm N}^+$ end of the ${\rm P}^{-}-{\rm N}^+$ dipole) in the
screening of the electrostatic repulsion between the DNA chains. The
length and time scales of atomistic simulations are limited by memory
and CPU requirement to several nanometers and nanoseconds, which is
far below the macroscopic regime encompassing the statistics and
evolution of large molecular ensembles. At the macroscopic level, only
the continuum behavior of existing CL-DNA structures can be addressed
based on free energy functionals which are insensitive to the fine
details of the lipids and DNA
\cite{harries_may_benshaul,may_benshaul,bruinsma}. Electrostatic
screening effects between the DNA chains resulting from {\em
non-specific}\/ interactions between the lipids and DNA have been
reported in these studies. This observation is complementary to the
specific mechanisms observed in detailed atomistic computer models.

In this paper we present an ``intermediate'' molecular modeling approach that
retains the most essential components of self-assembly and molecular
statistics, but avoids the computational overhead of a full atomistic
model. The model \cite{farago_jensen_pincus} extends an existing
coarse-grained (CG) molecular bilayer model
\cite{farago,farago_pincus} by including both charged and neutral
lipids, as well as charged DNA molecules. The inter-molecular
potentials between different molecular species are designed to mimic
the hydrophobic effect {\em without}\/ the explicit presence of
solvent. Thus, the approach carefully balances the need for molecular
detail with computational practicality in a manner that allows for
solvent-free simulations of complex self-assembly over long enough
time scales to address experimental reality. In addition to showing
spontaneous self-assembly of CL-DNA complexes, we also investigate the
structural properties of lamellar complexes and measure a number of
important quantities such as the dependence of the interaxial distance
between DNA chains on the fraction of charged lipids, the polarization
of the cationic charge distribution, local area density fluctuations,
and the effective two-dimensional stretching elastic modulus,
$K^*_A$. Some of these quantities are not easily accessible to
theoretical continuum models and may require further
approximations. One of the more interesting results is the decrease in
$K^*_A$ upon increasing the membrane charge density $\sigma_M$, which
reflects reduced mechanical stability and a higher probability of
structural defects, such as membrane pores. The observations of such
pores is consistent with the experimental findings of enhanced
transfection efficiencies at high concentration of CLs, because pores
must be formed to enable the escape of the DNA from the complex (see
discussion above). It demonstrates the utility of CG modeling in
addressing some key features of complex biological systems in general,
and lipid-DNA assemblies in particular.

We focus on isoelectric lamellar complexes in which the total charges
carried by the CLs and DNA are equal. The mechanism of counterion
release is most effective at this point, making the free energy of the
complex minimal \cite{harries_may_benshaul}. The counterions
concentrations inside the complex depend on their bulk
concentrations. Here, we study the very low bulk concentration limit
in which the counterions are (almost) completely depleted from the
complex. This regime has been previously addressed in atomistic
computer simulations of CL-DNA complexes
\cite{bandyopadhyay_terek_klein}. Counterions effects have been dealt
with in the framework of continuum Poisson-Boltzmann (PB) theory
\cite{harries_may_benshaul,may_benshaul,bruinsma}. While this approach
captures certain features of the charge distribution and the electric
fields in the complex, it neglects several important factors such as
the discrete nature of the ions and their finite sizes, which may
govern their distribution in the gaps of (sub)nanometer dimensions
that exist between the membranes and the DNA. Treating the water in
these tiny voids as a bulk medium of effective dielectric constant
$\epsilon=78$ and neglecting dehydration effects are other gross
approximations made by most current modeling techniques.

We use a continuum model to analyze some of the simulation
results. Our analytical study is based on the minimization of a
phenomenological free energy functional with respect to the profile of
the membrane and the cationic charge distribution. This free energy
includes contributions from all the electrostatic interactions
existing between the lipids and the (infinite array of) DNA molecules,
as well as terms associated with the mixing entropy and (small length
scale) protrusion modes \cite{lipowsky_grotehans} of lipids. We show,
both computationally and analytically, that the cationic charge
distribution is polarized. The minimum of the charge density is
obtained halfway between adjacent DNA molecules while the maximum is
not reached right above the DNA, but is slightly shifted. The origin
of this shift is the ability of the lipids not located right above the
DNA to protrude and thus position their charged headgroups in regions
of the complex where the electrostatic potential created by the DNA
array is lower. Interestingly, we find that the charge density in the
immediate vicinity of the DNA tends to match effective area charge
density of the DNA rod. Charge matching between narrowly separated
surfaces has been observed in previous studies of CL-DNA complexes
\cite{harries_may_benshaul,harries_may_benshaul2}, as well as in
studies of other molecular assemblies \cite{haleva_bental_diamant}. It
has been attributed to the increased concentration of ions which are
bound to remain in the confined volume between surfaces in order to
neutralize the system. Charge matching is favorable because it enables
the release of these strongly confined ions. Our study, which is based
on an ion-free model, suggests that charge matching can be also driven
by other mechanisms.

The paper is organized as follows: In the next section we present our
CG model of CL-DNA complexes and provide most of the details of the
simulations (except for the details of a new Monte Carlo (MC) scheme
that we use to sample the constant tension ensemble, which we
introduce in Appendix \ref{appendix:surfacetension}). The results of our
simulations are described in the two following sections dealing,
respectively, with the dependence of the DNA spacing and the
complex stability on the charge density, and the charge density
fluctuations. The computational results are compared to the
predictions of a analytical continuum model in Appendix
\ref{appendix:analytical}. We close the paper with a brief discussion
of the main results and an outline of some future prospects.

\section{Computer Model}

The computer model of CL-DNA complex is based on a bilayer CG model
presented elsewhere \cite{farago,farago_pincus}. The lipids are
modeled as short trimer molecules consisting of one ``hydrophilic''
and two ``hydrophobic'' beads which are connected to each other by
stiff linear springs \cite{farago_pincus}. The model does not include
explicit solvent. Rather, a set of short-range attractive
intermolecular potentials is used, which effectively mimic
hydrophobicity and allow self-assembly of bilayer from molecular
disorder \cite{farago_jensen_pincus}. Depending on the area density of
the lipids, the bilayer is found in either a solid or a fluid phase
\cite{farago}, where the latter is characterized by an in-plane lipid
diffusion and out-of-plane fluctuations whose spectrum is well
depicted by Helfrich effective Hamiltonian \cite{helfrich}. The model
of the CL-DNA complex is obtained by: (1) Choosing a fraction $\phi_c$
of the lipids and placing a unit (point) charge $+e$ at the centers of
their hydrophilic bead. (2) Introducing DNA molecules, each of which
is modeled as a rigid rod with a uniform axial charge density
$\lambda_{\rm DNA}=-e/1.7$\AA~and radius $R_{\rm DNA}=10$\AA. The
diameter of the spherical particles constituting the lipids is set to
$\sigma\simeq 6.3$\AA~(see definition of $\sigma$ in
Ref.~\cite{farago}), which yields an area per lipid $a_{\rm
lipid}\simeq 70$\AA$^2$ for uncharged bilayers. Excluded volume
interactions between rods (R) and spheres (S) are introduced via a
truncated (at $r_c=\frac{\sigma}{2}+R_{\rm DNA}$)
and shifted potential of the form:
$U_{RS}/k_BT=50\left\{\left[\left(\sigma/2+R_{\rm
DNA}\right)/r\right]^{12} -1\right\}$, where $r$ is the distance
between the center of the sphere and the axis of symmetry of the rod.
The distance between nearest-neighbor rods is restricted to $d_{\rm
DNA}\geq 2R_{\rm DNA}$.

Modeling the DNA strands as infinite rods carrying uniform charge
density $\lambda$ is consistent with the CG approach of the model,
where only electrostatics, noise, and simple geometric features are
retained. In this representation of DNA molecules we ignore the effects 
associated with their (1) flexibility and (2) the discrete nature of their 
charge distribution. The first approximation is justified in view of the 
fact that the DNA persistence length ($\xi_p\sim 500\,\AA$) is an order of
magnitude larger than all the other relevant length scales in the
problem. Curvature fluctuations involve free energy penalty of about
$1\,k_BT$ per $\xi_p$ of DNA length, which is negligible compared to
the complex stabilization free energy of $\sim 10^2-10^3\,k_BT$ per
persistence length \cite{harries_may_benshaul}. The second
approximation is supported by numerical studies revealing that the
electrostatic potential around the DNA surface is not much different
from that produced by the continuous charge density, except for a
narrow regime in its immediate vicinity \cite{wagner}.

We study isoelectric complexes where the total charges of the DNA and
the CLs neutralize each other, with no added counterions. Simulations
of the quasi two-dimensional (2D) complex are conducted in a
rectangular system of size $L_x\times L_y\times L_z$, with full
periodic boundaries along the $x$ and $y$ directions, and periodicity
with respect to only lipid mobility and short-range interactions in
the $z$ direction. The simulations were performed at room temperature
and with a bulk water uniform dielectric constant $\epsilon=78$. The
rods are arranged in a 1D array, parallel to the $y$ axis and with
equal spacing along the $x$-direction. Long range electrostatic
interactions between the charged spheres, infinite rods, and their
periodic images were accounted for using the Lekner summation method
\cite{niels}. The electrostatic potential energy, per simulation cell,
between a CL whose charged headgroup is located at
$\vec{r}+\Delta\vec{ r}\equiv(x+\Delta x,y+\Delta y,z+\Delta z)$ and
another CL and its replicas located at $(x+mL_x,y+nL_y,z)$,
where $m,n$ are integers, is given by the exponentially convergent
summation of, e.g., the form:
\begin{eqnarray}
V_{SS}\left(\Delta\vec{r}\right)
&=&\frac{e^2}{\epsilon}\left\{\frac{4}{L_x}\sum_{n=1}^{\infty}\cos\left
(2\pi\frac{\Delta x}{L_x}n\right)\right.\nonumber \\
&\times&
\sum_{k=-\infty}^{\infty}K_0
\Biggl[2\pi n \sqrt{\left(\frac{L_y}{L_x}\right)^2\left(\frac{\Delta y}{L_y}
+k\right)^2+
\left(\frac{\Delta z}{L_x}\right)^2}\,\Biggr]\nonumber \\
&-&\left.\frac{1}{L_x}\ln\left[\cosh\left(2\pi\frac{\Delta z}{L_y}\right)
-\cos\left(2\pi\frac{\Delta y}{L_y}\right)\right]^{2}-\frac{\ln 2}{L_x}
\right\},
\end{eqnarray}
where $K_0$ is the modified Bessel function of order $0$. The
self-energy $V_{SS}^{(0)}$ that arises from a charged sphere with its
own periodic images is found by evaluating the expression
\begin{equation}
V_{SS}^{(0)}=\frac{1}{2}\lim_{|\Delta\vec{ r}|\rightarrow
0}\left(V_{SS}\left(\Delta\vec{r}\right)-
\frac{e^2}{\epsilon|\Delta\vec{r}|}\right)
\end{equation}
and is given in Ref.~\cite{niels}. The sphere-rod electrostatic
energy per simulation cell is the combined logarithmic interactions between
a point charge and a 1D array of line charges. It is given by \cite{niels2}
\begin{equation}
V_{RS}\left(\Delta\vec{r}\right)=-\frac{e\lambda}{\epsilon}\ln\left\{
2\left[\cosh\left(2\pi\frac{\Delta z}{L_x}\right)-\cos\left(2\pi\frac{\Delta x}
{L_x}\right)\right]\right\}.
\label{eq:sphererod}
\end{equation}
The rod-rod electrostatic self-energy is \cite{niels2}
\begin{equation}
V_{RR}^{(0)}=-\frac{\lambda^2L_y}{\epsilon}\ln\frac{2\pi}{L_x}.
\end{equation}

Simulations of electrostatics in water-free models are usually
conducted with a uniform dielectric constant due to the complexity of
including mirror charges in a disordered molecular ensemble. In PB
theories \cite{harries_may_benshaul,haleva_bental_diamant} a different
type of boundary conditions (BCs) is usually assumed, namely that the
dielectric constant vanishes in the interior of the DNA and the lipid
membrane. This latter approximation is justified by the fact that the
(bulk) water dielectric constant, $\epsilon=78$, is much larger than
all the other relevant dielectric constants. The electric field lines
prefer to stay in high dielectric media and the limit $\epsilon=0$
corresponds to systems in which the electric field is entirely
contained within the aqueous part. Fortunately, the exclusion of
electric fields from the DNA and the membranes is not solely related
to their low dielectric constants. In our simulations, it is mainly a
matter of the geometry and the charge distribution in the
system. Therefore, the electrostatic forces, through which the charges
in water interact with each other, are expected to be insensitive to
the dielectric constants of the DNA and the membranes. Computational
studies \cite{niels2} of electrostatics near similar simple geometric
interfaces indicate that the net effect of mirror images is, indeed,
minor. This observation also serves as a justification for our choice
of BCs. Rather than using full periodic BCs in all three directions,
we study an infinite ``slab'' consisting of a single array of charged
DNA molecules and two bilayers (see Fig.~\ref{fig:complex1}), in which
only the inner monolayers (facing the DNA rods) are charged while the
outer monolayers consist of neutral lipids only. The interaction
between different slabs (whose overall net charge is neutral) is well
screened and is significantly weaker than the Coulomb interactions
between the charged components within each slab.

In a preceding publication \cite{farago_jensen_pincus}, we have shown
that complexes such as the one appearing in Fig.~\ref{fig:complex1}
are formed spontaneously in simulations starting from a disordered
initial state where the lipids are randomly distributed within the
simulation cell. This demonstrates that the complex represents a
stable equilibrium phase of the system. In this work we focus on
structural properties of CL-DNA complexes and use pre-assembled
complexes for this purpose. The simulations were performed at constant
surface tension $\gamma=0$ by employing a new sampling scheme to
generate area-changing trial moves. The sampling scheme, which is
different from the commonly used method of sampling the $(N,\gamma,T)$
ensemble \cite{groot_rabone}, is described in detail in Appendix
\ref{appendix:surfacetension}. The rest of the details of the
simulations appear in Ref.~\cite{farago_jensen_pincus}.

\section{DNA spacing and mechanical stability}

Measuring the DNA interaxial spacing, $d_{\rm DNA}$, serves as a
critical test to our model's ability to mimic the mesoscale behavior
of CL-DNA complexes, because $d_{\rm DNA}$ can be measured in X-ray
diffraction experiments. The experimental data of Safinya et al.
\cite{safinya_science1,safinya_currop} shows that for isoelectric
complexes, the dependence of $d_{\rm DNA}$ on the membrane charge
density $\sigma_M=2e\phi_c/a_{\rm lipid}$ is governed by the
relationship \cite{safinya_iso}
\begin{equation}
d_{\rm DNA}=\frac{\lambda_{DNA}}{\sigma_M}=
\left(\frac{a_{\rm lipid}\lambda_{\rm DNA}}{Ne}\right)\frac{1}{\phi_c},
\label{eq:ddna}
\end{equation}
which results from simple mass conservation in the lamellar
geometry. 

The computational results for the average spacing between adjacent DNA
rods, $d_{\rm DNA}$, are plotted in Fig.~\ref{fig:ddna} as a function
of the inverse of the fraction of charged lipids, $1/\phi_c$.  The
solid line is a fit to Eq.~(\ref{eq:ddna}) with $a_{\rm
lipid}=69$\AA$^2$ which is the area per lipid in uncharged
membranes. The deviation from linear behavior at high charge densities
arises from the increase in $a_{\rm lipid}$ with $\phi_c$ (see
Fig.~\ref{fig:lipidarea}). The numerical data is in excellent
agreement with the experimental results reported in
Ref.~\cite{safinya_science1}. Specifically, the experimental $d_{\rm
DNA}$ vs.~$\phi_c$ data [see Fig.~4 (B) in
Ref.~\cite{safinya_science1}] show agreement with Eq.~(\ref{eq:ddna})
at low charge densities (with a value of $a_{\rm lipid}$ which is
slightly different than the one defined here) and a similar deviation
trend at high charge densities (note that our Eq.~(\ref{eq:ddna}) and
the comparable one in Ref.\cite{safinya_science1} express the same
relationship in different forms. See discussion in
\cite{safinya_iso}.) The assumption underlying Eq.~(\ref{eq:ddna}) is
that the effective interactions between the DNA are repulsive and
balanced by the elastic membrane forces. Linear elastic stress acting
on a membrane is related to $a_{\rm lipid}$ and its equilibrium value
$a^0_{\rm lipid}$ by $\tau=K_A\left(a_{\rm lipid}-a^0_{\rm
lipid}\right)/a_{\rm lipid}^0$, where $K_A$ is the 2D stretching
modulus, which for lipid bilayers is typically in the range
$K_A\gtrsim 10^2\ {\rm ergs}/{\rm cm}^2$. At high charge densities,
the electrostatic stress is sufficiently large to eliminate the
membrane thermal undulations and increase $a_{\rm lipid}$
\cite{lau_pincus}. In the present study, we find for complexes with
$\phi_c\sim0.85$ that the strain $\varepsilon\equiv\left(a_{\rm
lipid}-a^0_{\rm lipid}\right)/a_{\rm lipid}^0\sim 0.1$, which is
somewhat larger than the typical strain lipid bilayers withstand
before rupture ($\varepsilon\sim0.02-0.05$, \cite{boal}). The
discrepancy between experimentally observed rupture strains and the
strain found in our model is not surprising given the coarse grained
lipid-model of simple three point objects. It may also be partially
attributed to the system size dependence of the rupture strain
\cite{tolpekina}. Membranes with higher $\phi_c$
have indeed been found to be susceptible to pore formation, as
illustrated by the configuration in Fig.~\ref{fig:porecomplex} of a
complex with $\phi_c\sim 0.9$. The loss of mechanical stability is
also evident from the rapid decrease in the effective stretching
modulus of the complex $K_A^*$ for $\phi_c\gtrsim 0.7$
(Fig.~\ref{fig:modulus}), which has been extracted from the mean
square of fluctuations in $a_{\rm lipid}$: $K_A^*=k_BT\,a_{\rm
lipid}^0/\left[N\left\langle\left(a_{\rm lipid}-a_{\rm
lipid}^0\right)^2\right\rangle\right]$. The larger area fluctuations
at high $\phi_c$ increase the probability of pore opening which in
turn may lead to disassociation of the complex.

It is interesting to compare the inter-DNA electrostatic interactions
in the complex with the same interactions in bulk in the presence of
monovalent counterions \cite{podgornik1}. In both cases, the bare
electrostatic repulsion is screened by the distribution of microions
and CLs around the DNA. In the complex, the distribution of the
cationic charge is limited by a geometric constraint, namely the
residence of the CLs on the membrane surface which maintains a finite
separation from the inter-DNA plane. Therefore, we expect that
screening to be less efficient than in bulk solutions. It is also
reasonable to expect that, compared to 3D systems, the confinement to a
surface increases the repulsive electrostatic interactions between the
cationic charges which further increases the magnitude of the
effective inter-DNA repulsive force. 

Since our model does not include water explicitly, the hydration
forces which may be significant at small $d_{\rm DNA}$ are missing
from the picture \cite{podgornik2}. This can be corrected either by
introducing an additional short range DNA-DNA potential that
explicitly account for the hydration free energy or assuming that
$R_{\rm DNA}$ represents the hydrated rather than the bare DNA
radius. Since the hydration forces decay on DNA-DNA surface
separations of more then $\sim$ 1 nm, their introduction into the
model will only modify the results at small $d_{\rm DNA}$ (large
$\phi_c$). Specifically, this will increase the effective DNA-DNA
repulsion and, therefore, will strengthen the trend observed in
Fig.~\ref{fig:ddna} that at high charge densities Eq.~(\ref{eq:ddna})
underestimates $d_{\rm DNA}$. It will also shift the limit of
mechanical stability to slightly lower charge densities. Another
feature missing in our model is the contribution of DNA bending
fluctuations to the inter-DNA interaction. The effect of these
fluctuations in bulk is to increase the decay length of both hydration
and electrostatic forces \cite{podgornik3}. One may expect a similar
contribution to inter-DNA interactions in CL-DNA complexes, although
we are unaware of any systematic study of this effect in 2D. X-ray
studies find weak positional disorder in CL-DNA complexes. The typical
correlated domain size of the 1D lattice of DNA extends to nearly 10
unit cells \cite{safinya_science1}, which is twice as large than the size
of the complex in our simulations.

Our computational results explain well the recently observed enhanced
TE of lamellar CL-DNA complexes at high charge densities
\cite{safinya_currmed,safinya_charge}. The limiting stage of the
transfection process is the escape of the complex from the endosome in
which it is initially trapped after entering the cell. Escape from
the endosome occurs through activated fusion of the complex and
endosomal membranes, during which both must be perforated. Having a
complex with poor mechanical stability is an advantage at this stage,
since such a complex will tend to open pores more easily, and through
these pores the DNA may be released to the cytoplasm. We suggest that
the loss of mechanical stability results from the cationic charge of
the lipids and the pressure which it exerts on the complex
membrane. At high charge densities this pressure exceeds the rupture
tension of the membrane and, thus, leads to mechanical failure of the
complex.

\section{Charge density modulations}

The quantity defined as $\phi_c$ represents the {\em mean}\/ number
fraction of charge lipids. However, the lateral distribution of
cationic charge on the membranes need not be uniform. One may expect
the CLs to accumulate above and below the negatively charged DNA
rods. This tendency to minimize the electrostatic energy of the
CLs-DNA interactions is competed by the thermally induced
mixing entropy and the
repulsive electrostatic interactions between the CLs, which favor
homogeneous composition of the cationic and neutral
lipids. Furthermore, charge density modulations may be coupled to
membrane undulations \cite{harries_may_benshaul2,schiessel_aranda},
and both can contribute to lowering the free energy of the complex. As
discussed above, the stability of the complex is directly related to
its TE and, therefore, it is important to study the effect of these
``degrees of freedom'' of the lipids.

The dependence of $\phi_c$ on $x$, the position within a unit cell of
the complex (i.e., the interval between adjacent DNA rods) is depicted
in Fig.~\ref{fig:fraction}, where $x=0$ and $x=d_{\rm DNA}$ correspond
to lipids located right above or below the DNA (see inset of
Fig.~\ref{fig:fraction}). The curves, from bottom to top, are for the
following value of the mean number fraction:
$\phi_c=150/500=0.3;150/375=0.4;150/300=0.5;150/250=0.6;150/220\sim0.68;150/195\sim0.77;150/185\sim0.81$. As
expected, we find that for all values of $\phi_c$, the minimum of
$\phi_c(x)$ is achieved for $x=d_{\rm DNA}/2$, i.e., in the middle of
the unit cell. The minimum is more pronounced for low values of
$\phi_c$, in which case the maximum of $\phi_c(x)$ is at $x=0$ and
$x=d_{\rm DNA}$. At the higher values of $\phi_c$, the maximum shifts
from the edge of the unit cell towards the center and, in general, the
fluctuations in $\phi_c(x)$ become quite small.

The shift in the maximum of $\phi_c(x)$ from the immediate vicinity of
the DNA towards the center of the cell has been previously reported in
theoretical studies of the system based on PB theory
\cite{harries_may_benshaul,harries_may_benshaul2}. It has been
attributed to the tendency of the system to match the areal charge
density of the membrane with the effective areal charge density of the
DNA, $\sigma_{\rm DNA}\equiv -\lambda_{\rm DNA}/(2\pi R_{\rm DNA})\sim\
9.4\cdot10^{-3} e/$\AA$^2$. This involves attraction of CLs towards the
DNA at low values of $\phi_c$ and significant charge modulation over the
relatively large distance between the DNA rods. On the other hand,
when $\phi_c$ is large and $d_{\rm DNA}$ is small, the charge density
fluctuations are weak and CLs must be depleted from above/below the
DNA to match the local charge density of the DNA.

The tendency to match the local charge densities of the membranes with
$\sigma_{\rm DNA}$ is seen more clearly in
Fig.~\ref{fig:chargedensity}. Here, the charge density $\sigma_M(x)$
rather than $\phi_c(x)$ is plotted as a function of $x$. The dashed
horizontal line corresponds to the effective charge density of the DNA
$\sigma_{\rm DNA}\sim 9.4\cdot10^{-3}e/$\AA$^2$. The graphs show that
for all values of $\phi_c$, the local charge density at the edge of
the unit cell remains within $15\%$ of $\sigma_{\rm DNA}$. More
interestingly, the graphs show that in contrast to $\phi_c(x)$, the
maximum of $\sigma(x)$ is always shifted from the DNA towards the
center of the unit cell. This observation deserves some special
comment: In most analytical studies of membranes, area density
fluctuations are neglected (or, rather, it is assumed that the local
area per lipid $a_{\rm lipid}(x)$ is constant) and, therefore,
$\phi_c(x)$ and $\sigma_M(x)=2e\phi_c(x)/a_{\rm lipid}(x)$ are
proportional to each other. Our computational results indicate that
area density fluctuations (see Fig.~\ref{fig:areadensity}) may be
quite important and serve as an additional degree of freedom that
further reduces the free energy of the system. In
Fig.~\ref{fig:areadensity}, $\rho_0=(a^0_{\rm
lipid})^{-1}=1/69$\AA$^{-2}$ denotes the area density of uncharged
membranes. For large $\phi_c$, $\rho<\rho_0$ for all values of $x$,
reflecting the fact the {\em mean}\/ area per lipid increases at high
charge densities (see Fig.~\ref{fig:lipidarea} and discussion
above). More noticeable are the area density fluctuations within the
unit cell which can be observed for all values of $\phi_c$. The
location of the maximum area density coincides with that of the
maximum charge density and, therefore, can be attributed to the
accumulation of charged lipids. Had the area density been constant,
this would mean depletion of the neutral lipids from the same region
in the unit cell. Area density fluctuations represent an additional
degree of freedom of the system, which permit a more uniform
distribution of the neutral lipids, and, thus, pays off in terms of
lower mixing entropy. The magnitude of the area density fluctuations
is roughly given by $\sqrt{(\rho/\rho_0-1)^2}\sim (k_BT/K_A^* a_{\rm
lipid}^0)^{1/2}$, which for typical values of the parameters
($K_A^*\sim 250\ {\rm ergs}/{\rm cm}^2$, $a_{\rm lipid}^0\sim
70$\AA$^2$ - see Figs.~\ref{fig:lipidarea} and \ref{fig:modulus})
yields $\sqrt{(\rho/\rho_0-1)^2}\sim 0.15$, in reasonable agreement
with our results in Fig.~\ref{fig:areadensity}.

As mentioned above, local charge density matching is the origin of the
CLs tendency to migrate towards the middle of the unit cell. Solutions
of the PB equation \cite{harries_may_benshaul,haleva_bental_diamant}
show that the concentration of counterions, which will be bound in the
narrow water gap that exists between the DNA and the membrane, increase with
the charge density mismatch. An accumulation of
counterions in such a small volume is energetically unfavorable and
will lead to a very large osmotic pressure. Two comments should
be made regarding the solution of the PB equation: First, the
screening of the electrostatic interactions by the highly confined
counterions is probably overestimated by continuum PB theory
because the slithering of counterions into the small gaps will be
hindered both dynamically (excluded volume) and thermodynamically
(dehydration). Second, our simulations of isoelectric complexes with
no counterions apply to the no-screening limit where this effect is
not expected to occur. We therefore conclude that local charge density
matching may be also driven by other factors. Indeed, in the absence
of screening one must consider the interactions of the CLs with the
periodic array of line charges rather than with the closest DNA
rod. The Coulomb energy due to the interaction of a charge $+e$ with
an infinite periodic array of rods of charge density per unit length
$\lambda<0$ is given by Eq.~(\ref{eq:sphererod}), with $L_x=d_{\rm
DNA}$. For a charge residing on a perfectly flat surface located a
distance $\Delta z=D\equiv R_{\rm DNA}+\sigma/2\sim 13$\AA~above the
mid-plane of the DNA array, Eq.~(\ref{eq:sphererod}) simplifies to
$V_{RS}\left(\Delta\vec{r}\right)\sim
\left(e\lambda/\epsilon\right)\, B\cos\left(2\pi\Delta x/d_{\rm DNA}\right)$,
with $B=2\exp\left(-2\pi D/d_{\rm DNA}\right)$. This expression is
valid as
long as $B<1$, which is indeed the case for the above value of $D$ and
the range of values of the DNA interaxial spacing $d_{\rm DNA}\sim
25$\AA$-50$\AA~considered in this work. For a nearly flat surface with
$\Delta z=D-h$ ($0\leq h\ll D$), the sphere-rod electrostatic energy
reads
\begin{equation}
V_{RS}\left(\Delta\vec{r}\right)\simeq \frac{e\lambda}{\epsilon}
\left[\frac{2\pi h}{d_{\rm DNA}}+B\cos\left(\frac{2\pi\Delta x}
{d_{\rm DNA}}\right)
\right].
\label{eq:sphererod2}
\end{equation}
The first term in this equation reflects the long-range nature of
unscreened electrostatic interactions which makes the potential of the
infinite array of line charges look similar to the potential of a
uniformly charged surface with areal charge density
$\sigma_M=\lambda/d_{\rm DNA}$. The second term is due to the
periodicity of the system and represents the tendency of cationic
lipids to favor the proximity of the anionic DNA rods ($\Delta
x=0$). The attraction of the CLs to the DNA rods, located at the edge
of the unit cell, will be offset by their attraction towards the {\em
mid-plane}\/ of the DNA array [first terms in
Eq.~(\ref{eq:sphererod2})]. This attraction draws the CLs towards the
center of the unit cell because, right above the DNA rod, the vertical
separation between the surface and the DNA array is restricted to
$\Delta z=D$ ($h=0$) by excluded volume interactions. Away from the
DNA and close the center of the unit cell, the elastic deformation of
the surface permits $\Delta z<D$ ($h>0$) which is energetically
favorable. The highest cationic charge density will be obtained at the
minimum value of the sphere-rod electrostatic energy $V_{RS}$. A
detailed calculation based on a continuum expression for the free
energy which includes contributions of the electrostatic and elastic
energies and of mixing entropy, is presented in Appendix
\ref{appendix:analytical}. We show that only for infinitely rigid
surfaces the maximum of charge density is observed at the edge of the
unit cell. In all other cases, the membrane tends to deform towards
the mid-plane of the DNA array, which leads to shifting of the maximum
of $\sigma_M(x)$ towards the center of the cell. This is in agreement
with our results in Fig.~\ref{fig:chargedensity}, although the
deformation of the membrane is unnoticeable in snapshots of the system
(e.g., Fig.~\ref{fig:complex1}). Our calculation shows that the
typical amplitude of the deformation is extremely small, of the order
of $1-2$\AA. This estimate is model dependent, but in agreement with
previous studies of the system
\cite{harries_may_benshaul2,schiessel_aranda}, and explains the
apparent flatness of the membrane observed in our simulations.

\section{Discussion}

We have presented a molecular simulation method which captures the
self-assembly of cationic liposomes complexed with DNA - a promising
synthetically based nonviral carrier of DNA for gene therapy. The
method is an intermediate modeling approach between atomistic computer
simulations and continuum phenomenological theories. Like the former,
it utilizes a molecular description of the system; but similarly to
the latter, it employs a coarse-grained (CG) representation of the
intra-molecular atomic details. The reduced number of degrees of
freedom, as well as the fact that the model does not require explicit
representation of the embedding solvent, lead to a significant improvement in
computational efficiency. Thus, the approach carefully balances the need
for molecular detail with computational practicality in a manner that
allows for solvent-free simulations of complex self-assembly over long
enough time scales to address experimental reality.

In addition to showing spontaneous self-assembly of cationic lipid-DNA
complexes, the broad utility of the model is illustrated by
demonstrating excellent agreement with X-ray diffraction experimental
data for the dependence of the interaxial distance between DNA chains,
$d_{\rm DNA}$, on the fraction of charged lipids
$\phi_c$. Specifically, we find that $d_{\rm DNA}$ is inversely
proportional to $\phi_c$ - a relationship which can be also derived by
a simple packing argument where the DNA rods form a space-filling 1D
lattice. This result is indicative of a repulsive long-range inter-DNA
interaction. The predominant contribution to this interaction is due
to {\em non-specific}\/ electrostatic repulsion between the negatively
charged DNA rods, which is only partially screened by the cationic
charge on the membranes. We note that the magnitude of the repulsive
interaction plays no role in the packing argument. Therefore, a linear
relationship between $d_{\rm DNA}$ and $\phi_c^{-1}$ is predicted by
both our CG simulations and the PB theory, despite of the fact that in
the PB treatment the screening of inter-DNA repulsion is also due to
the counterions presented in the complex. 

Certain features of CL-DNA complexes, for instance the process of
self-assembly and structural defects (Fig.~\ref{fig:porecomplex}), can
be addressed more effectively through CG simulations rather than by
continuum theories of existing structures. This point is nicely
demonstrated by our simulations of highly-charged complexes. Upon
increasing the fraction of the CLs, we find that: (1) the area per
lipid increases, (2) the effective stretching modulus of the complex
decreases, and (3) the rate of pore formation increases which
eventually leads to the disintegration of the complex. All together,
these results indicate that the higher the charge density of the
membranes the lower the mechanical stability of the system. This is a
key observation which may explain the recently observed enhanced
transfection efficiency (TE) of lamellar CL-DNA complexes at high
charge densities. Transfection is viewed as a two-stage process: (1)
cellular uptake via endocytosis, and (2) escape of the complex from
the endosome, presumably through fusion of the lipids with the
endosomal membrane and release of the DNA into the cytoplasm. TE of
lamellar complexes is limited by the rate of the second stage and,
hence, increases with the decrease of mechanical stability, i.e., with
increase of charge density.

Given the consistency of agreement between our CG molecular approach
and observed experimental features, we suggest that the presented
model is an appropriate and promising tool for investigating the
statistics and dynamics of lipid-DNA complexes on spatial and temporal
scales relevant for biological and biomedical applications. In future
work, we plan to develop models which would mimic CL-DNA complexes
with improved gene delivery performance, such as complexes containing
multivalent lipids and lipids attached to short polymer chains. A
special effort will be made to develop a model for the inverted
hexagonal (${\rm H_{II}^{\ C}}$) structure, and to examine the
mechanical behavior of this phase, which appears to be experimentally
quite distinct from the behavior of the lamellar phase. We will
also investigate the effect of counterions which must be presented in
the positively charged complexes that adhere to the negatively charged
cell membrane (at the initial stage of the transfection process). The
model may be also extended to include some features of the DNA helical
structure. These more advanced models may lead to a better
understanding of the principles governing the statistical-mechanical
behavior of CL-DNA complexes, which is crucial for systematic and
successful design of efficient synthetic vectors for gene therapy.

\appendix
\renewcommand\thesection{\arabic{section}}
\section{Simulations at constant surface tension}
\label{appendix:surfacetension}

Bilayer Membranes may be considered as narrow interfaces consisting of
lipids and the hydration layers that separate two aqueous bulk
phases. Simulations of liquid/liquid interfaces can be performed in a
variety of statistical ensembles. Assuming that the temperature $T$
and number of particles $N$ are fixed, one may use the $(N,T,V,A_p)$
ensemble, in which the total volume of the system, $V$, and the
projected area of the interface, $A_p$, are held
constant. Alternatively, the normal and transverse components of the
pressure tensor, $P_n$ and $P_t$ may be fixed, letting $V$ and $A_p$
fluctuate. Another common choice is the constant surface tension
ensemble $(N,T,V,\gamma)$, which mimics the experimental conditions
more closely than the $(N,T,V,A_p)$ ensemble does. The $(N,T,V,\gamma)$ is of
particular importance for simulations of membranes, which can exhibit
large undulations at vanishing surface tension. Accessing the
$\gamma=0$ regime is crucial to modeling such systems.

There has been an ongoing theoretical debate concerning the
statistical thermodynamic definition of $\gamma$. In analytical
studies the surface tension is usually regarded as the thermodynamic
variable conjugate to the total interface area $A$, which is the sum
of the projected area $A_p$, and the area stored in the thermal
undulations $\Delta A$. Because of the relatively high value of their
stretching modulus, bilayer membranes are often assumed to have a
fixed total area, and $\gamma$ is used as a Lagrange multiplier fixing
the value of $A$. However, in computer simulations it is difficult to
sample an ensemble where $A$ is constant since $\Delta A$ cannot be
easily controlled and, moreover, its value is not well defined (in
contrast to continuum models). It is, therefore, more common in
computer simulations that the surface tension is treated as conjugate
to $A_p$, which is the cross-sectional area of the simulation
cell. With this convention, one readily derives the following
relationship between $\gamma$ and the pressure tensor
\cite{rowlinson_widom}
\begin{equation}
\gamma=\left\langle L_n\times\left(P_n-P_t\right)\right\rangle,
\label{eq:tension1}
\end{equation}
where $L_n$ is the size of the system in the direction perpendicular
to the interface, and $\langle\rangle$ denotes thermal average. This
quantity coincides with yet another quantity commonly referred to as
the ``surface tension'', namely the $q^2$ coefficient in the
expression describing the dependence of the mean thermal fluctuations
on the wave vector (the ``spectral intensity'') \cite{farago_pincus}:
\begin{equation}
\left\langle |h_q|^2\right \rangle=\frac{k_BT}
{a_{\rm lipid}\left[\gamma q^2+\kappa q^4 + {\cal O}(q^6)\right]}.
\end{equation}

To simulate the $(N,T,V,\gamma)$ ensemble, one needs to sample
configurations, in which the total volume of the system is conserved
while the area is allowed to fluctuate. The common method to generate
such an ensemble is to consider a rectangular simulation box of volume
$V=L_x\times L_y\times L_z$ and projected area $A_p=L_x\times L_y$
and, occasionally, rescale the dimensions of the box and the molecular
coordinates $\{\vec{r}^{\,\, i}\}$ in the following manner
\cite{groot_rabone}:
\begin{eqnarray}
&\ &L_x \rightarrow L_x+\delta L_x\ ;\ r_x^i \rightarrow 
r_x^i\left(\frac{L_x+\delta L_x}{L_x}\right) \nonumber \\
&\ &L_y \rightarrow L_y+\delta L_y\ ;\ r_y^i \rightarrow 
r_x^i\left(\frac{L_y+\delta L_y}{L_y}\right) \nonumber \\
&\ &L_z \rightarrow \frac{V}{(L_x+\delta L_x)(L_y+\delta L_y)}\ ;\ r_z^i 
\rightarrow r_z^i\left[\frac{L_xL_y}{(L_x+\delta L_x)(L_y+\delta L_y)}\right].
\label{eq:sampling1}
\end{eqnarray}
In scaled coordinates,
$\{\vec{l}^{\,\,i}=(r^i_x/L_x,r^i_y/L_y,r^i_z/L_z)\}$, the partition
function is given by
\begin{equation}
Z=\int_0^{\infty} d\!L_x\,d\!L_y\,d\!L_z\,\delta
\left(L_z-\frac{V}{L_xL_y}\right)
\cdot \int_0^1 \Pi_{i=1}^{N} dl^i_x\,dl^i_y\,
dl^i_z\,\, V^Ne^{-\beta\gamma A_p}\,
e^{-\beta U\left(\{\vec{l}^{\,\,i}\},L_x,L_y,L_z\right)},
\label{eq:partition1}
\end{equation}
where $\delta$ is the Dirac delta function, $U$ is the energy of the
configuration and $\beta=1/k_BT$. Since the volume $V$ and number of
particles $N$ are both fixed, the acceptance criterion is given by
\begin{equation}
P_{\rm acc}(o\rightarrow n)=\min\left[1,e^{-\beta
\left(\delta U+\gamma\delta A_p\right)}\right],
\end{equation}
where $\delta U$ and $\delta A_p$ denote, respectively, the difference
in the energy and area between the new ($n$) and old ($o$)
configurations.

Eq.~(\ref{eq:sampling1}) defines a one-to-one, {\em locally}\/ volume
preserving, transformation of the molecular coordinates between
rectangular simulation boxes of slightly different dimensions. In
principle, however, only the {\em total}\/ volume of the simulation
cell must be conserved and, therefore, other one-to-one transformation
may be proposed. One such transformation, which is particularly
suitable for solvent-free interfacial systems, is the following:
\begin{eqnarray}
&&L_x \rightarrow L_x+\delta L_x\ ;\ r_x^i \rightarrow 
r_x^i\left(\frac{L_x+\delta L_x}{L_x}\right) \nonumber \\
&&L_y \rightarrow L_y+\delta L_y\ ;\ r_y^i \rightarrow 
r_x^i\left(\frac{L_y+\delta L_y}{L_y}\right)  \\
&&L_z \rightarrow \frac{V}{(L_x+\delta L_x)(L_y+\delta L_y)}\ ;\ r_z^i 
\rightarrow \left\{\begin{array}{ll}
r_z^i\left[\frac{L_xL_y}{(L_x+\delta L_x)(L_y+\delta L_y)}\right]\equiv
r_z^i+\delta r_z^i & i=1 \nonumber \\
{\rm mod}(r_z^i+\delta r_z^1,L_z(0)) & i\neq 1 
\end{array}
\right.,
\label{eq:sampling2}
\end{eqnarray}
where $L_z(0)$ is the initial (at $t=0$) height of the simulation box,
and ``mod'' is the modulus operator. Unlike transformation
(\ref{eq:sampling1}), only the coordinates of one particle [labeled
``1'' in Eq.(\ref{eq:sampling2})] are rescaled proportionally to the
size of the box in (\ref{eq:sampling2}). Therefore, {\it a-priory},
only this particle is guaranteed to remain within the rescaled
simulation box, while all the others may be displaced beyond the
boundaries of the system in the $z$ direction ($0\leq
z<L_z$). Detailed balance requires that in the case of such an event,
the move attempt will be rejected, i.e., the energy of the new
configuration is defined $U=\infty$. Let us consider an interface
located close to the center of the box ($z\sim L_z/2$), separating two
dense bulk liquid phases. Let us also assume that the particle with
index ``1'' resides on the interface. For such a solvent-containing
system, rescaling the dimensions of the simulation box and the
molecular coordinates according to transformation (\ref{eq:sampling2})
will usually fail. Decreasing $L_z$ [$L_z<L_z(0)$] will lead to
ejection of some solvent particles from the system, while increasing
$L_z$ [$L_z>L_z(0)$] will lead to the formation of an empty stripe and
changing the bulk densities which is energetically very costly. In
solvent-free models, on the other hand, the bulk phases are ``empty''
and such a problem will arise only if the interface is located close
to one of the $z$ boundaries. For infinitely long runs, the fraction
of time that the interface spends near the boundaries scales like
$w/L_z$ (where $w$ is the physical width of the interface), which can
be made arbitrarily small by increasing $L_z$ (i.e., by changing the
``volumes'' of the empty bulk phases). In practice, $L_z$ need not be
very large since the diffusion of the interface is so vanishingly slow
that it will never reach one of the boundaries within conceivable
simulation time. With function (\ref{eq:sampling2}) defining the
transformation between simulation boxes of different shapes, and in
terms of the scaled coordinates
$\{\vec{l}^{\,\,1}=(r^1_x/L_x,r^1_y/L_y,r^1_z/L_z);\vec{l}^{\,\,i\neq
1}=(r^i_x/L_x,r^i_y/L_y,r^i_z/L_z(0))\}$, the partition function in
(\ref{eq:partition1}) is rewritten:
\begin{eqnarray}
Z&=&\int_0^{\infty} d\!L_x\,d\!L_y\,d\!L_z\,\delta
\left(L_z-\frac{V}{L_xL_y}\right)\nonumber \\
&\cdot& \int_0^1 \Pi_{i=1}^{N} dl^i_x\,dl^i_y\,
dl^i_z\,\, V\left[L_z(0)A_p\right]^{N-1}\,e^{-\beta\gamma A_p}\,
e^{-\beta U\left(\{\vec{l}^{\,\,i}\},L_x,L_y,L_z,L_z(0)\right)},
\label{eq:partition2}
\end{eqnarray}
from which we readily conclude that the acceptance criterion should
be
\begin{equation}
P_{\rm acc}(o\rightarrow n)=\min\left[1,
\left(1+\frac{\delta A_p}{A_p}\right)^{N-1}e^{-\beta
\left(\delta U+\gamma\delta A_p\right)}\right].
\end{equation}
This criterion resembles the acceptance criterion for simulations of
the 2D isobaric-isothermal ensemble, accept for the exponent of the
term $\left(1+\delta A_p/A_p\right)$ being $N-1$ rather than $N$. This
could be interpreted in the following way: Simulating a thin interface
of width $w$ at constant $\gamma$ is similar to simulating a 2D
system at constant pressure. The center of mass of the interface can
be found with equal probability at any position along the $z$
direction of the simulation box. However, because of the total volume
conservation, the height of the (mostly empty) simulation box scales
with the inverse of the area $A_p$. The contribution of this degree
of freedom is expressed by the additional factor $\left(1+\delta
A_p/A_p\right)^{-1}$.

We have performed simulations of identical systems (both neutral
membranes and CL-DNA complexes) using sampling methods
(\ref{eq:sampling1}) and (\ref{eq:sampling2}), and measured the
probability distributions of the energy and the area of the
system. Both methods produced identical distribution functions, but
the time it took to obtain reasonably accurate results was
considerably shorter with the new sampling scheme
[Eq.~(\ref{eq:sampling2})] (about $2\cdot10^6$ MC time units) than with
the conventional one [Eq.~(\ref{eq:sampling1})] (roughly $10^7$ MC time
units). The superior effectiveness of the new scheme should be
attributed to larger area changes per reshaping attempt, $\delta A_p$,
that the new scheme permits, which were typically half an order
of magnitude larger than in the old scheme. We believe that the origin
of this is the fact that the membranes in our simulations are
``softer'' with respect to area changes than they are with respect to
the variations of their width. In the conventional sampling scheme, area
and width fluctuations are coupled by local volume conservation, which
greatly reduces the magnitude of acceptable reshaping moves. In the new
scheme this coupling is removed, enabling area variations which do not
simultaneously squeeze the layers against each other or pull them apart. 

\section{Analytical modeling}
\label{appendix:analytical}

In order to estimate the deformation of the membranes and the charge
density fluctuations within a unit cell of the complex, we consider
the system shown schematically in Fig.~\ref{fig:complex2}. This system
consists of three charge distributions: an infinite array of equally
spaced rods with density per unit length $\lambda<0$, and two surfaces
with {\em mean}\/ charge density $\sigma_M>0$ per unit area
(representing the monolayers facing the DNA array on each side). The
vertical distance $\Delta z(x)$ of the surfaces from the mid-plane of
the DNA is $\Delta z=D\equiv R_{\rm DNA}+\sigma/2\sim 13$\AA~at the
edge of the unit cell ($x=0,d_{\rm DNA}$), and may be smaller ($\Delta
z=D-h<D$ ; $h(x)>0$) for $0<x<d_{\rm DNA}$ due to the
electrostatically induced deformation of the surfaces. We denote
the charge density fluctuations by $\delta \sigma(x)$, and consider the
limit of small fluctuations $|\delta\sigma|\ll\sigma_M$, as well as the
limit of small membrane deformation $|h|/d_{\rm DNA}\ll 1$. As a
reference state for energy calculations, we take a complex with
perfectly flat surfaces and no charge density fluctuations. Since the
complex is isoelectric, the mean charge density of the surfaces is
related to the DNA linear charge density by
\begin{equation}
2\sigma_M=-\frac{\lambda}{d_{\rm DNA}},
\label{eq:iso1}
\end{equation} 
while
\begin{equation}
\int_{0}^{d_{\rm DNA}} \left[\frac{1}{2}\sigma_M\left(\nabla h\left(x\right)
\right)^2+\delta\sigma\left(x\right)\right]dx\simeq
\int_{0}^{d_{\rm DNA}} \delta\sigma\left(x\right)dx=0.
\label{eq:iso2}
\end{equation}

For a fixed value of $d_{\rm DNA}$, the free energy of a unit cell of
the complex per unit length in the $y$ direction (parallel to the DNA
rods) consists of the following contributions:

1. The Coulomb energy of the interaction between the DNA rods and the
charged surfaces which is given by [see Eq.~(\ref{eq:sphererod2})]
\begin{equation}
F_1=2\int_0^{d_{\rm DNA}} \left[\sigma_M+\delta\sigma(x)\right]
\frac{\lambda}{\epsilon}
\left[\frac{2\pi h\left(x\right)}{d_{\rm DNA}}+B\cos\left(\frac{2\pi x}
{d_{\rm DNA}}\right)\right]
\left[1+\frac{1}{2}\left(\nabla h\left(x\right)
\right)^2\right]dx.
\label{eq:e1}
\end{equation}
The prefactor 2 in this expression is due to the
two surfaces interacting with the DNA rods. Treating $B$ in
Eq.~(\ref{eq:e1}) as a small parameter [$B=2\exp\left(-2\pi D/d_{\rm
DNA}\right)\sim0.12$, for $D\sim 13$\AA~and $d_{\rm DNA}\sim 30$\AA,
which is of the same order of magnitude as $(\delta\sigma/\sigma)$ and
$(h/d_{\rm DNA})$], we see that the leading term in the above
expression is 
\begin{equation}
F_1^1=2\int_0^{d_{\rm DNA}} \sigma_M
\frac{\lambda}{\epsilon}
\frac{2\pi h\left(x\right)}{d_{\rm DNA}}dx=
-\frac{4\sigma_M^2d_{\rm DNA}}{\epsilon}\int_0^{d_{\rm DNA}}
\frac{2\pi h\left(x\right)}{d_{\rm DNA}}dx.
\label{eq:e11}
\end{equation}
The superscript in $F_1^1$ (as well as in the other terms of the free
energy appearing below) denotes the order of the term in the small
parameters of the expansion. The next (second order) term is given by
\begin{eqnarray}
F_1^2&=&2\int_0^{d_{\rm DNA}} \delta\sigma(x)
\frac{\lambda}{\epsilon}
\left[\frac{2\pi h\left(x\right)}{d_{\rm DNA}}+B\cos\left(\frac{2\pi x}
{d_{\rm DNA}}\right)\right]dx\nonumber \\
&=&-\frac{4\sigma_Md_{\rm DNA}}{\epsilon}\int_0^{d_{\rm DNA}} \delta\sigma(x)
\left[\frac{2\pi h\left(x\right)}{d_{\rm DNA}}+B\cos\left(\frac{2\pi x}
{d_{\rm DNA}}\right)\right]dx.
\label{eq:e12}
\end{eqnarray}

2. The Coulomb energy of the interaction between the two charged surfaces,
which in leading order is given by
\begin{equation}
F_2^2=+2\frac{\sigma_Md_{\rm DNA}}{\epsilon}
\int_0^{d_{\rm DNA}} \delta\sigma(x)
\frac{2\pi h\left(x\right)}{d_{\rm DNA}}dx.
\label{eq:e22}
\end{equation}

3. The Coulomb energy of the interactions between lipids residing on
the same surface (the surfaces self electrostatic energy) [see
Eq.~(\ref{eq:sphererod}) with
$\lambda\rightarrow[\sigma_M+\delta\sigma(x')]dx'$ and $\Delta
z/L_x\rightarrow -h/d_{\rm DNA}\rightarrow 0$, and recall
Eq.~(\ref{eq:iso2})]
\begin{equation}
F_3^2=-\int_0^{d_{\rm DNA}} dx\left\{
\frac{\sigma_M^2d_{\rm DNA}}{\epsilon}G\left(\frac{h(x)}{d_{\rm DNA}}\right)
+\int_0^{d_{\rm DNA}} dx'
\frac{\delta\sigma(x)\delta\sigma(x')}{\epsilon}
\ln\left[2-2\cos\left(2\pi\frac{x-x'}{d_{\rm DNA}}\right)\right]\right\}
\label{eq:e32}
\end{equation}
where $G\sim \left(h/d_{\rm DNA}\right)^2$ is a dimensionless function.

4. The mixing entropy of the charged and neutral lipids in each
surface which, ignoring the variations in the area per lipid (see
Fig.~\ref{fig:areadensity}), is given by
\begin{equation}
F_4^2=\frac{k_BT}{a_{\rm lipid}}\int_0^{d_{\rm DNA}}
\frac{\delta\sigma^2\left(x\right)}
{\sigma_M\left(e/a_{\rm lipid}-\sigma_M\right)} dx ,
\label{eq:e42}
\end{equation}
where $e/a_{\rm lipid}$ is the maximum possible charge density
(obtained when all the lipids are charged) and, obviously,
$\sigma_M\leq e/a_{\rm lipid}$.

5. The elastic energy of the deformation of the surfaces. In previous
theoretical studies of CL-DNA complexes
\cite{harries_may_benshaul2,schiessel_aranda}, this energy has been
associated with the bending of the surfaces and has been expressed in
terms of Helfrich effective Hamiltonian \cite{helfrich}. However,
Helfrich effective Hamiltonian captures the elasticity of surfaces
only on length scale which are typically larger than the length of the
unit cell $d_{\rm DNA}\sim 25-50$\AA. At smaller scales, the elastic
behavior of membranes is dominated by individual or collective lipid
protrusions. Protrusion modes tend to increase the local surface area
and the restoring force acting against these protrusions can,
therefore, be characterized by an effective {\em local}\/ surface tension
$\gamma_p$. The corresponding elastic energy (per unit length in the
$y$ direction) is given (for two surfaces of length $d_{\rm DNA}$ in
the $x$ direction) by
\cite{lipowsky_grotehans}
\begin{equation}
F_5^1=\gamma_p \int_0^{d_{\rm DNA}}
\left(\frac{dh\left(x\right)}{dx}\right)^2 dx.
\label{eq:e51}
\end{equation}
The crossover from long scale bending-dominated to short scale surface
tension-dominated elasticity occurs on length scale $l$ which is of
the order of $l\sim 2\pi\sqrt{\kappa/\gamma_p}$, where $\kappa$ is the
bending modulus. Our previous measurements \cite{farago} give
$\kappa\sim 40k_BT$ and $l\sim 50$\AA, which yields $\gamma_p\sim
25\times10^{-2}\ {\rm J/m^2}$. Values of $\gamma_p$ measured for other
computer models of bilayer membranes
\cite{lindahl_edholm,marrink_mark} have been found to be of the same
order of magnitude - which happens to be comparable to the surface
tension of water-oil interfaces. Notice that the superscript ``1''
rather than ``2'' is used in Eq.~(\ref{eq:e51}), i.e., we consider this
term as linear in the expansion parameters despite of the fact
that the elastic energy appears as a second order term in $(h/d_{\rm
DNA})$. This is because the surface tension $\gamma_p$ is an
order of magnitude larger than the energy per unit area
$\sigma_M^2d_{\rm DNA}/\epsilon$ appearing in $F_1^1$ (\ref{eq:e11})
and, therefore, these two terms are comparable to each other. The
surface tension $\gamma_p$ is also an an order of magnitude larger
than the thermal energy per lipid area $k_BT/a_{\rm lipid}$, appearing
in the second order term $F_4^2$ (\ref{eq:e42}) which is associated
with the mixing entropy.

\begin{figure}
  {\centering \hspace{1.5cm}\epsfig{file=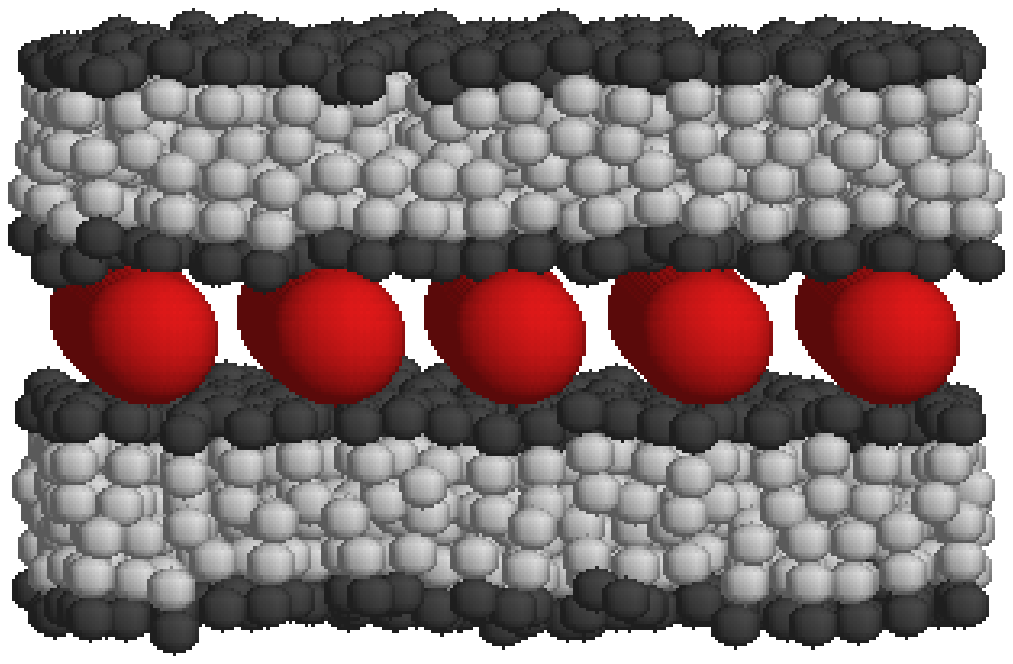,width=11cm}}

\vspace{3.0 in} 

\caption{Equilibrium configuration of a complex consisting of two bilayer 
    membranes, each with 390 lipids and five DNA strands. The lipids
    are modeled as trimers with hydrophilic (black) and hydrophobic
    (gray) particles. The DNA (red) are modeled as rigid
    rods with a uniform negative axial charge density. Then complex is
    isoelectric, i.e., the negative charge of the DNA is neutralized
    by the charge of the cationic lipids with no added salt. Thus,
    each bilayer in the shown complex includes 150 monovalent lipids,
    all of which reside in the inner layers facing the DNA array.}
\label{fig:complex1}
\end{figure}

\begin{figure}
  {\centering \hspace{1.5cm}\epsfig{file=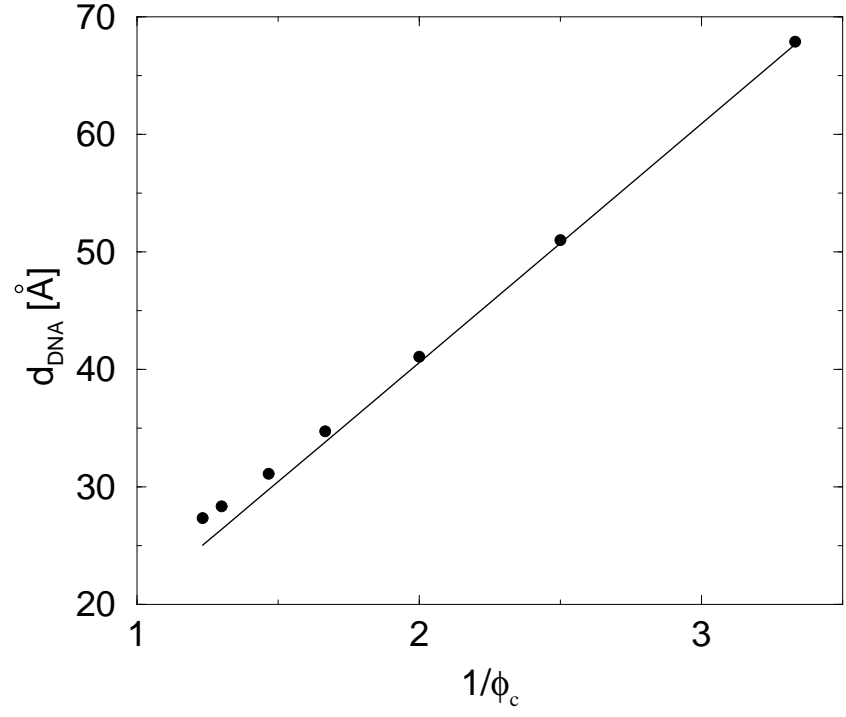,width=11cm}} 

\vspace{3.0 in} 

\caption{Average DNA spacing, $d_{\rm DNA}$, as a function of the
    inverse of the fraction of charged lipids $1/\phi_c$. Markers -
    numerical results (uncertainties are smaller than symbols); solid
    line - fit to Eq.~(\ref{eq:ddna}) with $a_{\rm lipid}=69$\AA$^2$.}
\label{fig:ddna}
\end{figure}

\begin{figure}
  {\centering \hspace{1.5cm}\epsfig{file=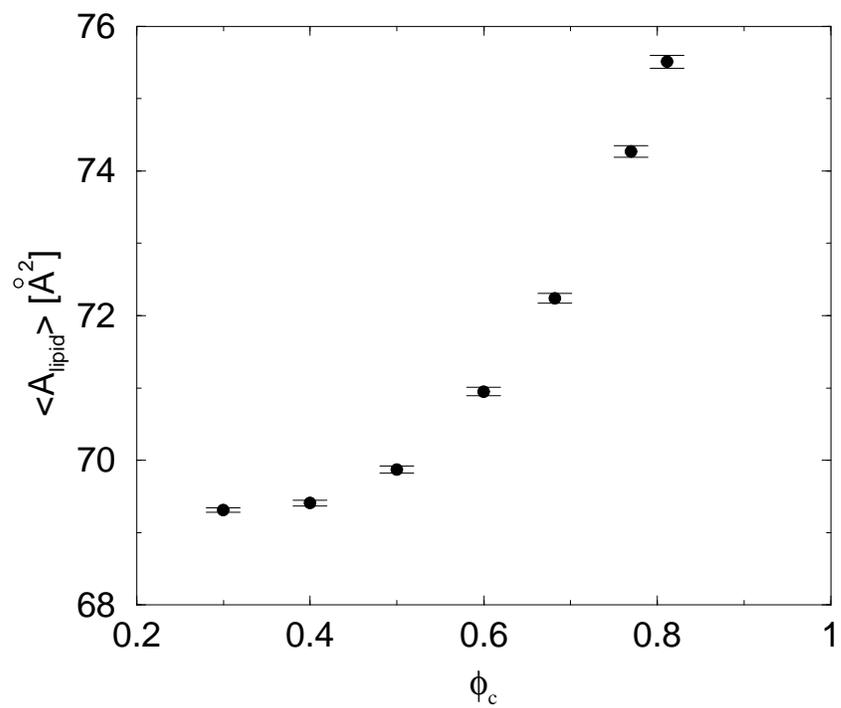,width=11cm}} 

\vspace{3.0 in} 

\caption{The average area per lipid, $a_{\rm lipid}$, as a function of
    the fraction of charged lipids $\phi_c$.}
\label{fig:lipidarea}
\end{figure}

\begin{figure}
  {\centering \hspace{1.5cm}\epsfig{file=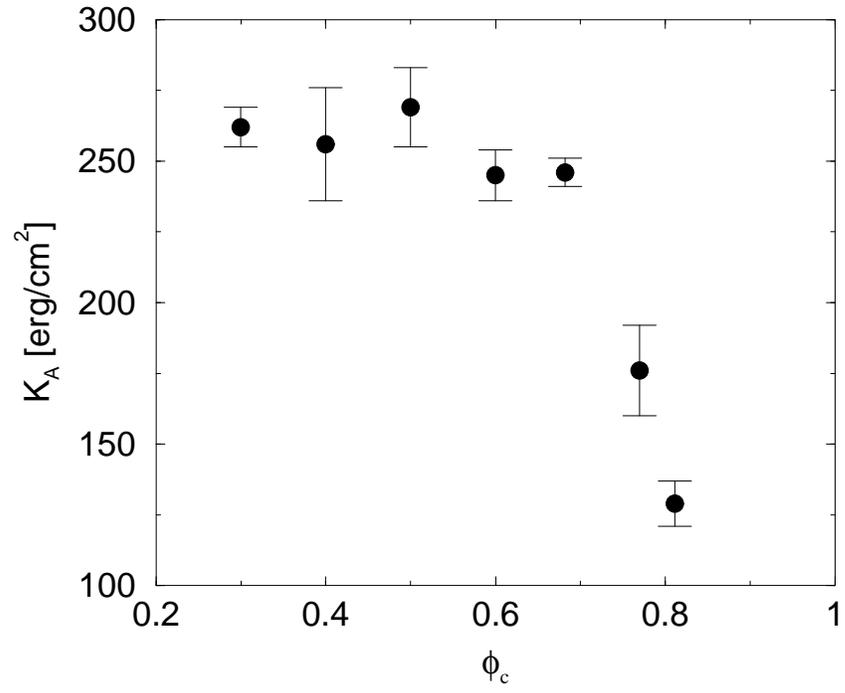,width=11cm}} 

\vspace{3.0 in} 

\caption{The effective stretching modulus, $K_A^*$, of the complex as a 
    function of $\phi_c$.}
\label{fig:modulus}
\end{figure}

\begin{figure}
  {\centering \hspace{1.5cm}\epsfig{file=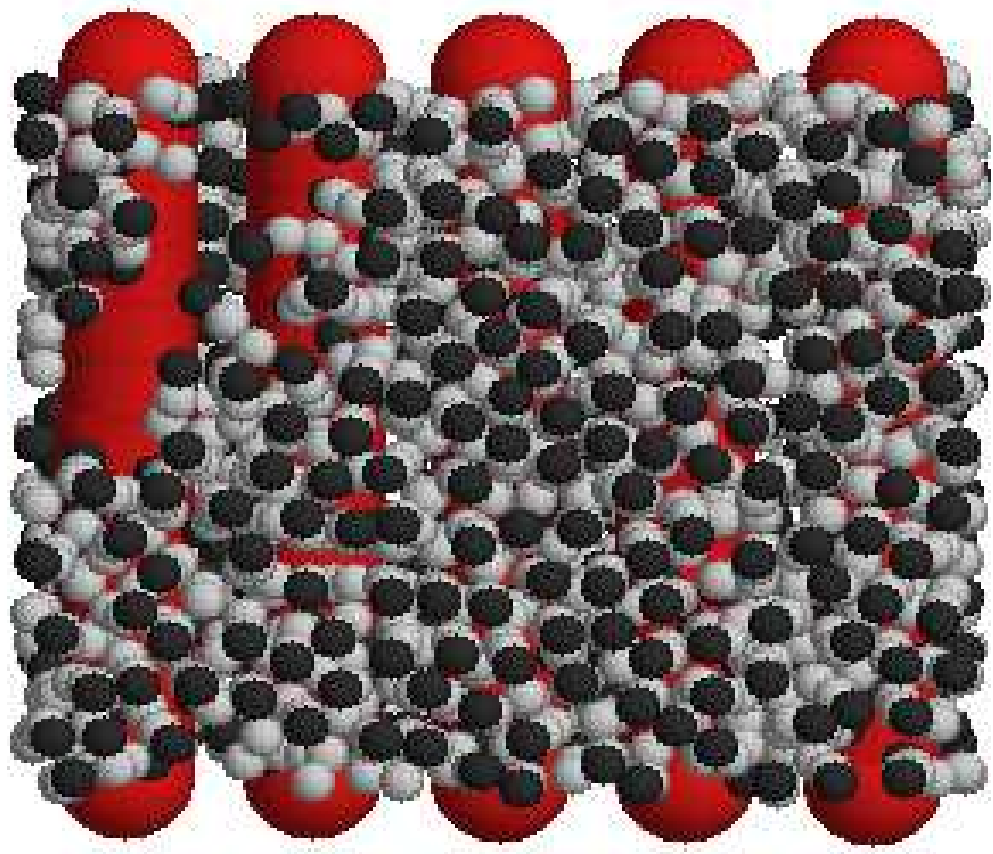,width=13cm}} 

\vspace{3.0 in} 

\caption{Equilibrium configuration of a complex with $\phi_c\sim0.9$
whose membranes develop pores.}
\label{fig:porecomplex}
\end{figure}

\begin{figure}
  {\centering \hspace{1.5cm}\epsfig{file=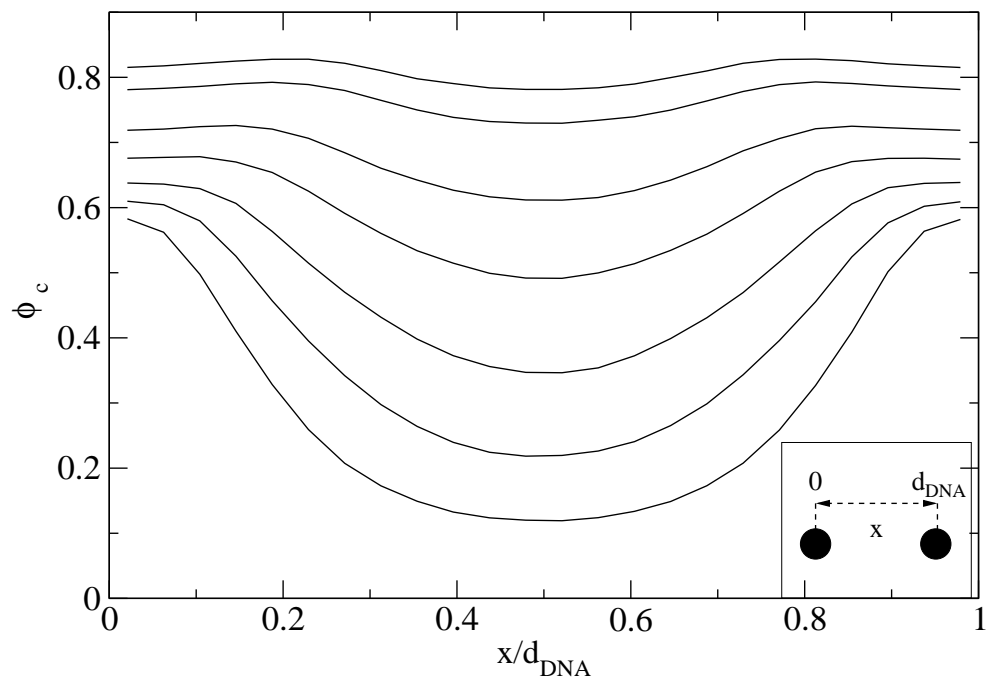,width=13cm}} 

\vspace{3.0 in} 

\caption{Local fraction of charged lipids $\phi_c$ as a function of $x$, 
    the position within a unit cell of the complex. Curves, from
    bottom to top, correspond to {\em mean}\/ fraction of
    0.3, 0.4, 0.5, 0.6, 0.68, 0.77, 0.81.}
\label{fig:fraction}
\end{figure}

With the above expressions for the various terms in the free energy of
the system, the equilibrium profile ${\tilde h}(x)$ of the surfaces is
found by minimizing the leading first order contribution
$F^1=F_1^1+F_5^1$. The function ${\tilde h}(x)$ is determined by the
Euler-Lagrange equation
\begin{equation}
\frac{d^2{\tilde h}(x)}{dx^2}+\xi^{-1}=0,
\label{eq:eulerlagrange}
\end{equation}
where the length scale is
$\xi\equiv\epsilon\gamma_p/(2\pi\sigma_M^2)$. The solution to
Eq.~(\ref{eq:eulerlagrange}), with boundary conditions
${\tilde h}(0)={\tilde h}(d_{\rm DNA})=0$, is
\begin{equation}
{\tilde h}(x)=\frac{1}{2\xi}x\left(d_{\rm DNA}-x\right).
\end{equation}
The maximum deformation of the surfaces is obtained at the center of
the unit cell ($x=d_{\rm DNA}/2$). For typical values of the physical
parameters in the system
\begin{equation}
a_{\rm lipid}\sim 70{\rm \AA}^2\ ;\ \sigma_M\sim0.8\frac{e}{a_{\rm lipid}}\ ;\ 
d_{\rm DNA}\sim 30{\rm \AA} ; \ \gamma_p\sim 25\times10^{-2}\ \frac{{\rm J}}
{{\rm m^2}},
\label{eq:typical}
\end{equation}
${\tilde h}(d_{\rm DNA}/2)\sim 1$\AA$\ll d_{\rm DNA}$, which justifies
treating $(h/d_{\rm DNA})$ as a small parameter and explains the
apparent flatness of the surfaces observed in the simulations (e.g.,
Fig.~\ref{fig:complex1}). It should be noted that our analysis
ignores the very small overlap that exists between the surface ${\tilde
h}(x)$ and the excluded volume of the DNA rods near the edges of the
unit cell.

\begin{figure}
  {\centering \hspace{1.5cm}\epsfig{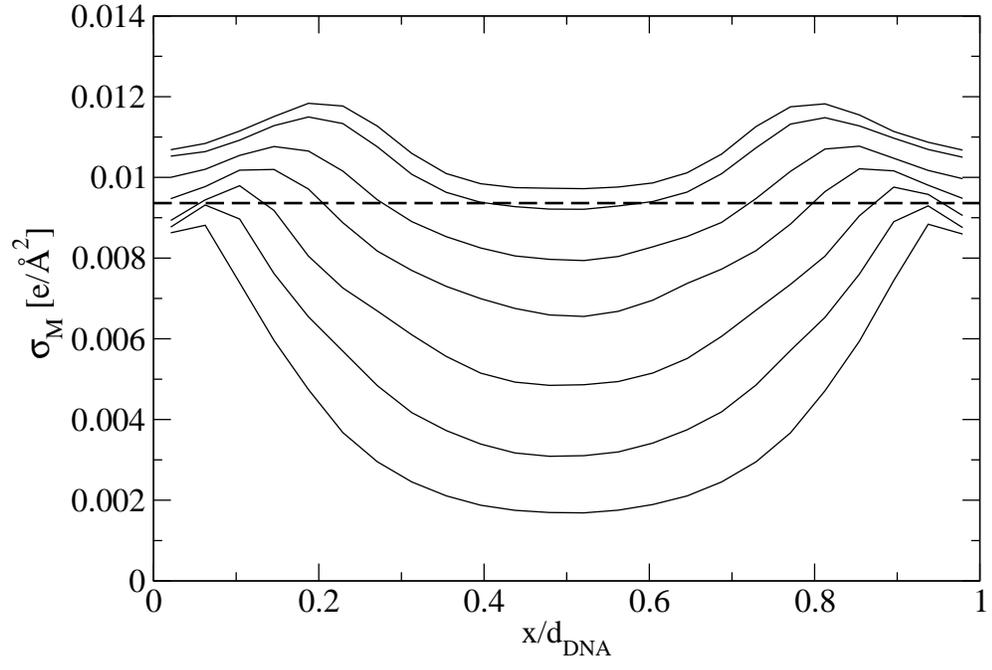}} 

\vspace{3.0 in} 

\caption{Local charge density of the membranes $\sigma_M$ as a 
    function of $x$. Curves, from bottom to top, correspond to
    $\phi_c=\,$0.3, 0.4, 0.5, 0.6, 0.68, 0.77, 0.81. Dashed
    horizontal line corresponds to the effective charge density of the
    DNA $\sigma_{\rm DNA}\sim 9.4\cdot10^{-3}e/$\AA$^2$.}
\label{fig:chargedensity}
\end{figure}

\begin{figure}
  {\centering \hspace{1.5cm}\epsfig{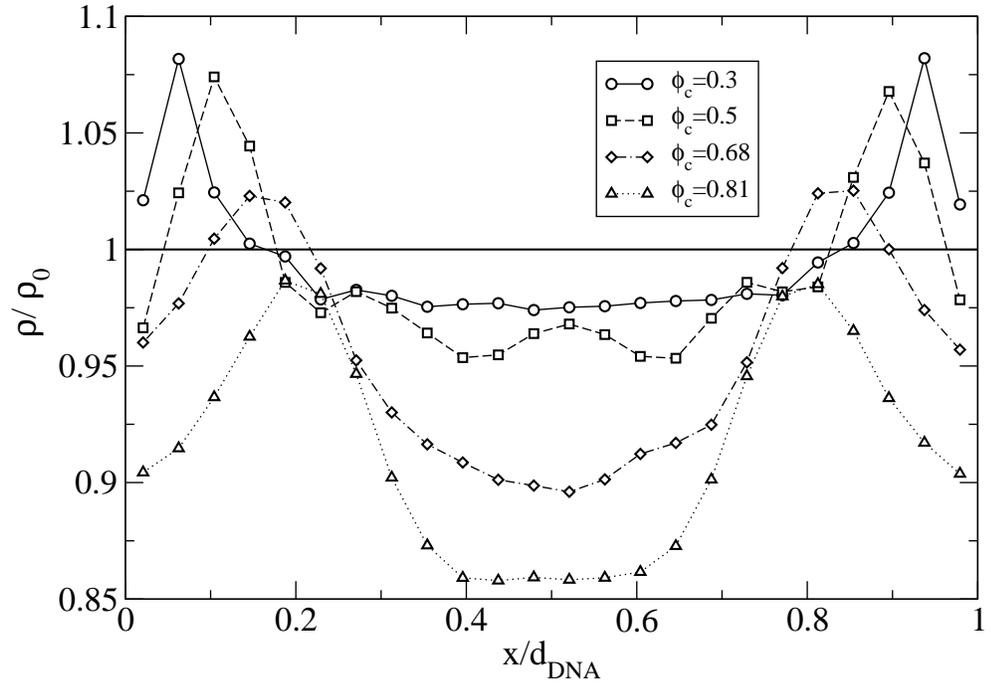}} 

\vspace{3.0 in} 

\caption{Total area density of the lipids $\rho$ as a function of 
    $x$ for different values of $\phi_c$. $\rho_0=(a^0_{\rm
    lipid})^{-1}=1/69$\AA$^{-2}$ is the area density of uncharged
    membranes. Lines are guide to the eyes.}
\label{fig:areadensity}
\end{figure}

\begin{figure}
  {\centering \hspace{1.5cm}\epsfig{file=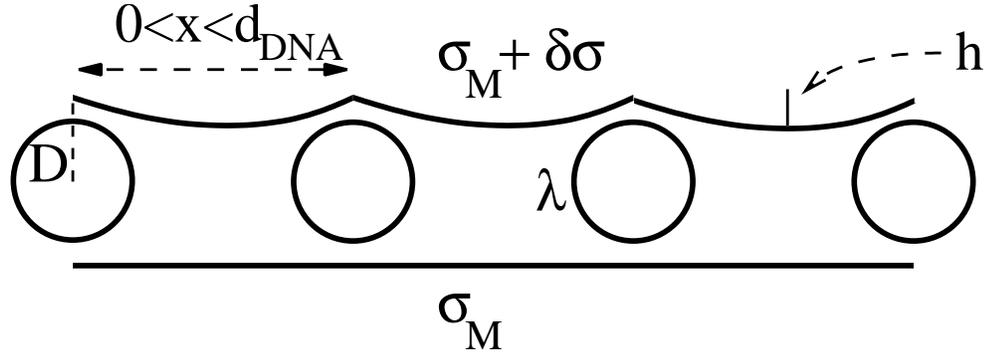,width=13cm}}

\vspace{3.0 in} 

\caption{Schematic picture of the complex consisting of an array
  of equally spaced DNA rods with nearest neighbor spacing $d_{\rm
  DNA}$ and two surfaces separated a distance $D$ from the mid-plane
  of the DNA array. The DNA rods are uniformly charged with charge
  density $\lambda<0$ per unit length. Surfaces have a mean charge
  density $\sigma_M>0$ per unit area and local charge density
  $\sigma_M+\delta\sigma(x)$. Their local height above/below the DNA
  mid-plane is denoted by $D-h(x)$. Lower surface is drawn in the
  reference state, where $\delta\sigma=0$, and, $h=0$.}
\label{fig:complex2} 
\end{figure}

The charge density fluctuations $\delta\sigma(x)$ will be determined by
minimizing the free energy given by the sum of second order terms
$F^2=F_1^2+F_2^2+F_3^2+F_4^2$, under the total charge conservation
constraint (\ref{eq:iso2}) and with $h(x)={\tilde h}(x)$. Expressing
$\delta\sigma(x)$ as a Fourier series
\begin{equation} 
\delta\sigma(x)=\sum_{n=-\infty}^{+\infty}
C_ne^{2\pi in\left(x/d_{\rm DNA}\right)},
\end{equation}
we find, after some algebra, that the optimal charge distribution is
obtained for the Fourier series
\begin{equation}
C_n=\frac{\sigma_Md_{\rm DNA}}{\epsilon}
\left[\frac{B\delta_{|n|,1}-d_{\rm DNA}/\left(2\xi n^2\right)}
{k_BT/\left(a_{\rm lipid}\sigma_M\left[e/a_{\rm lipid}-\sigma_M\right]\right)
+d_{\rm DNA}/\left(\epsilon |n|\right)}\right],
\end{equation}
where $\delta$ in the numerator denotes Kronecker's delta. An
approximate, but more useful, form for $\delta\sigma(x)$ can be
obtained by noting that typical values of the physical parameters
[see Eq.~(\ref{eq:typical})] lead to the first term in the denominator being
larger than the second term for all values of $n$. Thus, a reasonable
approximation may be obtained by dropping the smaller term, which
effectively means neglecting the term $F_3^2$ [Eq.~(\ref{eq:e32})] in
the second order free energy. Without this term, the real space form
of $\delta\sigma$ is given by
\begin{equation}
\delta\sigma(x)=
\sigma_M\frac{\sigma_Ma_{\rm lipid}}{e}\left(1-
\frac{\sigma_Ma_{\rm lipid}}{e}\right)
\frac{2d_{\rm DNA}l_B}{a_{\rm lipid}}B
\left\{\cos\left(\frac{2\pi x}{d_{\rm DNA}}\right)
+A\left[\frac{x\left(d_{\rm DNA}-x\right)}{d_{\rm DNA}^2}
-\frac{1}{6}
\right]\right\},
\label{eq:deltasigma}
\end{equation}
where $l_B=e^2/(\epsilon k_BT)\sim 7.1$\AA~is the Bjerum length, and
$A=B^{-1}(\pi/2)\cdot(d_{\rm DNA}/\xi)$. From this expression we
readily conclude the following: (1) Only for $A=0$, which corresponds
to the limit of infinitely rigid surfaces
($\gamma_p\rightarrow\infty$), is the peak of the charge distribution
at the edge of the unit cell. (2) Upon increasing $A$, the maximum of
$\delta\sigma(x)$ shifts gradually towards the center of the unit
cell, and (3) stays in the center, $x=d_{\rm DNA}/2$, for all values
$A\geq 2\pi^2$. One can easily verify that for the range of parameters
in our simulations, $0<A<2\pi^2$, which explains our observation of
the maximum of $\delta\sigma(x)$ somewhere between the edge and the
center of the unit cell (see Fig.~\ref{fig:chargedensity}). We can
also use Eq.~(\ref{eq:deltasigma}) to estimate the amplitude of the
charge fluctuations. For the physical parameters given in
Eq.~(\ref{eq:typical}), we have $B\sim 0.12$. The maximum of
$\delta\sigma(x)$ is obtained for $x\sim0.1d_{\rm DNA}$ (which should be
compared to $x\sim0.2d_{\rm DNA}$ in the simulations), where
$\delta\sigma/\sigma_M\sim 0.07$ (compare to
$\delta\sigma/\sigma_M\sim 0.1$ in the simulations). We consider this
semi-quantitative agreement with the numerical results as reasonable,
given the approximate nature of our analytical model.


\newpage

$\;$

\newpage       

$\;$

\newpage       

$\;$

\newpage       

$\;$

\newpage       

$\;$

\newpage       

$\;$

\newpage       

$\;$

\newpage       

$\;$

\newpage       

\end{document}